\documentclass[aps,superscriptaddress,a4paper,twocolumn]{revtex4}
\usepackage{blindtext}
\usepackage{graphicx}
\usepackage{amssymb}
\usepackage{color}
\usepackage{amsmath}
\usepackage{empheq}
\usepackage{lipsum}
\usepackage{physics}
\usepackage{mathrsfs}
\usepackage{MnSymbol}
\DeclareMathOperator{\arcsinh}{arcsinh}
\DeclareMathOperator{\arctanh}{arctanh}
\DeclareMathOperator{\Ci}{Ci}

\newcommand{\be}{\begin{equation}}
\newcommand{\ee}{\end{equation}}

\newcommand{\PR}[1]{\ensuremath{\left[#1\right]}}
\newcommand{\PC}[1]{\ensuremath{\left(#1\right)}}

\begin{document}
\title{Localized structures in two-field systems: exact solutions in the presence of Lorentz symmetry breaking and explicit connection with geometric constraints}

\author{G. H. Bandeira
\footnote{Email: ghbandeira111@gmail.com}}

\author{D. Bazeia
\footnote{Email: bazeia@fisica.ufpb.br}}

\author{G. S. Santiago
\footnote{Email: gss.santiago99@gmail.com}}
\affiliation{{Department of Physics, Federal University of Paraiba, João Pessoa, Paraíba, Brazil}}

\author{Ya. Shnir
\footnote{Email: shnir@theor.jinr.ru}}

\affiliation{{BLTP JINR, Joliot-Curie St 6, Dubna, Moscow region, 141980}}

\begin{abstract}
We investigate a class of models described by two real scalar fields in two-dimensional spacetime. The study focuses mainly on the presence of exact static solutions which satisfy the first-order formalism, in models constructed to engender Lorentz symmetry violation. We start by exploring a direct connection between Lorentz breaking and geometric constraint, as experimentally examined in the case of domain walls in geometrically constrained magnetic materials. By means of a specific choice of functions, we show that imposing geometric constraint within the Lorentz-violating framework recovers the exact solutions of the corresponding Lorentz-invariant theory. Furthermore, we extend the investigation to new models that go beyond reproducing the Lorentz invariant geometrically constrained solutions, revealing that it remains possible to parametrize the first-order equation of one of the fields through a suitably redefined coordinate. 
\end{abstract}

\maketitle

\section{Introduction}

Topological defects appear in systems whose vacuum manifold has multiple points or, mathematically speaking, where the homotopy group is nontrivial. These solutions acquire a structure relative to the scenario in which they are inserted, such as kinks, domain ribbons and domain walls for purely scalar field theories, vortices and monopoles for theories in which complex scalar fields are coupled to Abelian or non-Abelian gauge fields, respectively \cite{vi,ma,va,shnir}. These objects, endowed with an internal structure, appear in several areas of nonlinear science, particularly in field theory and cosmology \cite{vi,ma,va,shnir,Amen,dark}, and in condensed matter physics \cite{bi2,bi1,malomed}. Specifically, kinks are finite energy solutions of a real scalar field that interpolate between two distinct vacuum points, appearing in a two-dimensional spacetime \cite{ma}. When embedded in (3+1) spacetime they manifest as two-dimensional structures known as domain walls. Although kinks are among the simplest localized structures, they still yield significant results that can be used in applications in nonlinear science, in particular, in optical fibers and lattices, where the scalar field may be used to model the optical signal in the material, and also in Bose-Einstein condensates, where the scalar fields can describe the matter contents of the condensates. There are many works in these subjects, and here we suggest Refs. \cite{OF,BEC,Malo}, which offer a long and diverse list of theoretical and experimental investigations and references. Moreover, kinks are of direct importance in the study of magnetic structures in magnetic materials \cite{BOOKa,BOOK}. Since the scalar fields are agents that break the symmetry, they may be seen as order parameters, in a way similar to the spontaneous polarization which is investigated in ferroelectric materials. In this sense, the present study is also of interest in applications in ferroelectrics; see, for instance, Refs. \cite{Fe01,Fe02} and references therein. The point here is that ferroelectric materials may also engender negative capacitance, which is directly related to the double-well shape of the ferroelectric polarization–energy landscape, similar to the double-well potential present in the standard model for kinks in a relativistic scalar field theory.

Due to the simplicity and importance of kinks in high energy physics and in applications in other areas of nonlinear science, in this work we will be concerned with the presence of analytical kink-like configurations in models described by two real scalar fields. Although the subject has started long ago, we will bring novelty with the addition of Lorentz symmetry breaking and the presence of geometric constraints. As one knows, the breaking of Lorentz and CPT symmetries implies that particles and their respective antiparticles may not have uniquely defined characteristics; consequently, they might exhibit preferred directions in spacetime and may not possess the same charge; see Ref. \cite{Coste,Kos,Nature,Review,Petrov} and references therein for more on the subject. On the other hand, violation of relativistic invariance may lead to enhanced group of conformal symmetry \cite{Hagen:1972pd,Galushkina:2025hkw}. Lorentz violation is also present in Hořava-Lifshitz type theories; this appeared before in \cite{HL} and is of current interest in quantum gravity which was recently reviewed in \cite{QG}. In the case of real scalar fields, different models can be found, including the possibility to induce Lorentz symmetry breaking, as investigated, for instance, in Refs. \cite{lubo,barreto,baz01}. In \cite{lubo} the author considered a model that is generalized to include higher-order derivatives of the scalar fields, and in \cite{barreto} the Lorentz symmetry is broken via the presence of a constant vector, which is used to couple one scalar field to the derivative of the other. Moreover, in \cite{baz01} the Lorentz breaking is controlled by the presence of a second-order constant tensor, which mediates the coupling between the derivatives of the two scalar fields. In this work, we shall follow \cite{barreto}, so we further consider the Lorentz-breaking term with a constant vector in the next section.  

Before we start, it is worth to remind that one of the first field theory models that takes into account the Lorentz violation was proposed in \cite{Carroll}, with the addition of the term multiplied by a constant vector $k^{\mu}$ that introduces anisotropy in spacetime and thus breaks the Lorentz symmetry.
The addition of this term is interesting because of its application at both the classical and quantum level \cite{Petrov}. At the classical level, field models with Lorentz-violating extensions can offer a description of certain phenomena in condensed matter \cite{Grushin,Silva,Kosteleck},
since the classical solution behavior displays a role of an electromagnetic wave propagating in a medium with birefringence \cite{Jackiw,Schreck} and rotation of the polarization plane \cite{Jackiw}. See also Refs. \cite{LVA,LVB} for Lorentz violation in other contexts. Moreover, the anisotropy present under Lorentz violation is also of current interest in applications in condensed matter, as there are materials in which the breaking of isotropy plays important role for the appearance of localized structures \cite{BOOKa,BOOK,Fe01,Fe02}. A particularly interesting possibility is the Dzyaloshinskii-Moriya interaction \cite{Dzy,Mo}, which is known to produce effects to stabilize magnetic domain walls and skyrmions in magnetic systems, in particular, in the field of modern spintronics; see, e.g., Ref. \cite{Spin} and references therein. In connection with this, we also recall the interesting experimental investigation on magnetic material \cite{PRB}, which showed that the presence of a geometric constriction directly contributes to induce internal structure on the kink-like domain wall there investigated. This is an interesting issue, which we will further investigate in the present work, focusing on the connection between Lorentz breaking and the geometric constraint, as studied before in Ref. \cite{geometrical}. See also Refs. \cite{RA,RB,RC,RD} for investigations concerning the geometric constraint in other contexts of current interest.

In the present work one considers models described by two real scalar fields with kink-like solutions in a Lorentz-violating scenario. In Section \ref{ME} we develop the methodology that will be used throughout the work. In particular, we define the Lagrangian density and write the equations of motion and the corresponding energy-momentum tensor. We shall consider natural units $(\hbar=c=1)$ and assume that the fields, the spacetime coordinates, and all the parameters are rescaled in a way that leads us to consider the system described in terms of dimensionless quantities. We also deal with static fields and develop the first-order formalism for the theory, leading to analytical solutions and unveiling the connection between Lorentz breaking and geometric constraints, as previously explored in \cite{geometrical}, where kink-like solutions with internal structure were theoretically constructed and shown to reproduce the profiles observed for magnetic domain walls in geometrically constrained materials \cite{PRB}.
In Section \ref{IL} we use the results of the previous section to investigate distinct families of models in which the coupling of the fields only occurs via the Lorentz-breaking term. In the first family, a specific choice of functions was made so that we could reproduce the solutions of the Lorentz-invariant geometrically constrained model \cite{geometrical}. In the second family, we do not impose the need to replicate the aforementioned solutions, but we explore a simplified scenario where the auxiliary function depends only on one of the fields. However, in the third family, we analyze a more complex case in which the auxiliary function depends on both fields, which leads to interesting features, such as localized solutions with regions of negative energy density. The results presented in this work are all novel, and they unveil a direct and interesting connection between Lorentz symmetry breaking and geometric constraints, which may open another line of practical use concerning applications in condensed matter and in other areas of nonlinear science. We conclude the paper in Section \ref{Con}, where we offer some final remarks and outline different perspectives for future work.

\section{Methodology}
\label{ME}

The investigation described in the present work focuses on models containing two real scalar fields, where a coupling term between the two fields is used to explicitly break the Lorentz symmetry. The system is defined in (1 + 1) spacetime dimensions, with the metric chosen as $\eta_{\mu\nu} = \textrm{diag}(1,-1)$. In this case, the Lagrangian density is defined by
\begin{eqnarray}
\label{lgeo}
    \mathcal{L} = &&\frac{1}{2}\partial_{\mu}\phi\partial^{\mu}\phi + \frac{1}{2}\partial_{\mu}\chi\partial^{\mu}\chi+\nonumber\\ &&k^{\mu}g(\phi)\PC{1-\alpha f(\chi)}\partial_{\mu}\chi - V(\phi,\chi),
\end{eqnarray}
where $\alpha$ is a real parameter, and $g(\phi)$ and $f(\chi)$ are real functions of the fields. The derivative term $k^{\mu}g(\phi)\PC{1-\alpha f(\chi)}\partial_{\mu}\chi$, with $k^{\mu}=\PC{0,b}$ being a constant real vector, is included to introduce anisotropy in the system, which explicitly violates the Lorentz symmetry. If $\alpha = 0$ and $g(\phi)= \phi$, we restate the model investigated before in \cite{barreto}. So, in the present work we shall only consider the cases where $\alpha \neq 0$. Moreover, if $f(\chi) = \chi^2$ and $g(\phi) = \phi$ we retrieve the model studied in \cite{adam}, where the authors investigated the spectral wall phenomenon \cite{adamspe} for kink dynamics in a multi-field model engendering Lorentz violation and also, in the inspection of antikink-kink collisions in the $\phi^6$ model \cite{adam2}.  Note that investigation of localized solutions in 1+1 dimensional Lorentz invariant two-component scalar models revealed a lot of interesting features; see, e.g., Refs.\cite{Raj,Ru,GT,BB,BSR,BNRT,SV,Guilarte,Ya,Alonso,Were} and references therein for further details on the corresponding field configurations. Here we are mainly concerned with studying the influence of Lorentz breaking on the internal structure of the localized field configurations, an issue that may be of practical use in applications in condensed matter, related, for instance, to the magnetization in magnetic materials. As we demonstrate below, the choice of the Lorentz-breaking term is crucial to explicitly induce the geometric constraint. This is a novel phenomenon, which may also find applications in high energy physics, in particular, in braneworld models and cosmology. Regarding the importance of the formation of internal structures, an interesting possibility was reported before in \cite{PRB}, where a geometric constriction at the nanometric scale was able to significantly change the internal structure of the localized magnetic structure; see also Refs. \cite{BOOKa,BOOK} and references therein for more information on the subject. 

In the general case described by \eqref{lgeo}, the equations of motion are
\begin{subequations}\label{teom}
    \begin{align}
        \partial_{\mu}\partial^{\mu}\phi - k^{\mu}g_{\phi}\PC{1-\alpha f(\chi)}\partial_{\mu}\chi + V_{\phi} = 0, \\
        \partial_{\mu}\partial^{\mu}\chi + k^{\mu}g_{\phi}\PC{1-\alpha f(\chi)}\partial_{\mu}\phi + V_{\chi} = 0,
    \end{align}
\end{subequations}
where $V_{\phi} = \partial V/\partial \phi$ and $V_{\chi} = \partial V/\partial \chi$. Moreover, the energy-momentum tensor associated with this model takes the form
\begin{eqnarray}
\Theta^{\mu\nu}=&&\partial^{\mu}\phi\partial^{\nu}\phi + \partial^{\mu}\chi\partial^{\nu}\chi + \nonumber\\
&&k^{\mu}g(\phi)\PC{1-\alpha f(\chi)}\partial^{\nu}\chi - \eta^{\mu\nu}\mathcal{L}.
\end{eqnarray}
The corresponding energy density is
\begin{eqnarray}
\label{rhog}
    \rho = &&\frac{1}{2}\PC{\dot{\phi}^2 + \dot{\chi}^2 + \phi^{\prime2}\! + \chi^{\prime2}} -\nonumber\\
    &&bg(\phi)\PC{1\!-\alpha f(\chi)}\chi^{\prime} + V(\phi,\chi).
\end{eqnarray}
Since we will be searching for topological solutions, it is important to suppose boundary conditions that lead to localized solutions. In analogy with the standard theory, the boundary conditions considered within this work are such that $\phi(x=\pm\infty)$ and $\chi(x=\pm\infty)$ are minima of the potential, with $\phi^\prime(x=\pm\infty)=0$ and $\chi^\prime(x=\pm\infty)=0$. 

To facilitate the search for analytic solutions, it is useful to establish a
first-order framework \cite{bogo}, under which the equations of motion \eqref{teom} are satisfied by a set of coupled first-order differential equations. This is accomplished by introducing an auxiliary function $W =W(\phi,\chi)$ so that the energy density becomes
\begin{eqnarray}
    \rho =&& \!\frac{1}{2}\PC{\dot{\phi}^2 + \dot{\chi}^2} +\frac{1}{2}\PC{\phi' - W_{\phi}}^2 +\nonumber\\
    &&\frac{1}{2}\PC{\chi' - \PC{W_{\chi} + bg(\phi)(1-\alpha f(\chi))}}^2\!+ V-
    \nonumber\\
    &&\PC{\frac{1}{2}\!W_{\phi}^2\!+\! \frac{1}{2}\PC{W_{\chi}\!\!+\!bg(\phi)\PC{1\!-\alpha f(\chi)}}^2}\! +\! \frac{dW}{dx},\;\;
\end{eqnarray}
where $W_{\phi} = \partial W/\partial\phi$ and $W_{\chi} = \partial W/\partial\chi$. Considering static solutions, the first-order framework leads
\begin{subequations}
\label{1op}
 \begin{align}
 \dot{\phi} &= \dot{\chi} = 0,\\
  \phi' &= W_{\phi}, \label{primeira ordem de phi}\\
  \chi' &= W_{\chi} + bg(\phi)\PC{1-\alpha f(\chi)},\label{primeira ordem de chi}\\
  V(\phi,\chi) &= \frac{1}{2}W_{\phi}^2 + \frac{1}{2}\PC{W_{\chi} + bg(\phi)\PC{1-\alpha f(\chi)}}^2 \label{vg},
 \end{align}
\end{subequations}
where the first-order equations are only held for the positive sign and the energy is completely determined by the boundary conditions. Since the potential \eqref{vg} is nonnegative, this ensures that the pairs $(\phi(\pm\infty),\chi(\pm\infty))$ are global minima of the potential. Although these equations were obtained using a Bogomol'nyi-like procedure, we now search for positive energy solutions that satisfy the stressless condition, a property that also appears in models with generalized dynamics \cite{stress1,stress2}.
Notice that, due to the presence of a potentially negative term in the energy density \eqref{rhog}, the above procedure may not ensure that these solutions correspond to the lowest energy configurations compatible with the imposed boundary conditions.
Furthermore, it might be possible that the system of Eqs. \eqref{teom} also supports other non-BPS solutions, which we will not study in this paper.

\section{Illustration}
\label{IL}
\subsection{First Family}
The generalized theory previously developed exhibits an interesting property: choosing carefully the functions $g(\phi)$, $f(\chi)$ and $W(\phi, \chi)$, it is possible to find solutions that exactly match solutions of the geometrically constrained model studied before in Ref. \cite{geometrical}. To illustrate this feature, it is important to state the basic aspects of the theory whose solutions we aim to reproduce. As investigated in \cite{geometrical}, a geometric constraint emerges in a two real scalar field model when the fields are coupled through the introduction of a nonnegative function $h(\phi)$ in the kinetic term of $\chi$. To make a better correspondence with the theory we have developed so far, a modification of the nomenclature used in \cite{geometrical} is required. Specifically, we have to interchange $\phi$ with $\chi$ and replace $f(\chi)$ with $h(\phi)$ in \cite{geometrical}. The model is then described by the Lagrangian density
\begin{equation}
\label{lcons}
    \mathcal{L} = \frac{1}{2}h(\phi)\partial_{\mu}\chi\partial^{\mu}\chi + \frac{1}{2}\partial_{\mu}\phi\partial^{\mu}\phi - V(\phi,\chi).
\end{equation}
To establish a first-order framework, consider the energy density for static solutions
\begin{equation}
    \rho = \frac{1}{2}h(\phi)\chi^{\prime 2} + \frac{1}{2}\phi^{\prime 2} + V(\phi,\chi),
\end{equation}
and introduce an auxiliary function $\mathcal{W} =\mathcal{W}(\phi,\chi)$, such that
\begin{eqnarray}
    \rho= &&\frac{h(\phi)}{2}\PC{\chi'\mp \frac{\mathcal{W}_{\chi}}{h(\phi)}}^2 + \frac{1}{2}\PC{\phi'\mp \mathcal{W}_{\phi}}^2+ \nonumber\\
    &&\PC{V - \frac{1}{2}\frac{\mathcal{W}_{\chi}^2}{h(\phi)} - \frac{1}{2}\mathcal{W}_{\phi}^2} \pm \frac{d\mathcal{W}}{dx}.
\end{eqnarray}
With this, the first-order formalism gives the equations  

\begin{subequations}
    \begin{align}
        \chi' &= \pm \frac{\mathcal{W}_{\chi}}{h(\phi)}, \\
        \phi' &= \pm \mathcal{W}_{\phi},\\
        V(\phi,\chi) &=  \frac{1}{2}\mathcal{W}_{\phi}^2+\frac{1}{2}\frac{\mathcal{W}_{\chi}^2}{h(\phi)}.
    \end{align}
\end{subequations}

In order to reproduce the solutions of this model using the Lorentz-breaking framework introduced before, we have to establish a correspondence between the first-order  equations that appear in each one of the two models. Also, in order to explicitly express the geometric constraint it is considered that $W_{\chi} = 1-\alpha f(\chi)$. So, taking the upper sign in the above equations, the following restrictions must be satisfied

\begin{subequations}
\label{f2}
 \begin{align}
  W_{\phi} &= \mathcal{W}_{\phi}, \\
  W_{\chi} &= \mathcal{W}_{\chi},\\
   g(\phi) &= \frac{1}{b}\PC{\frac{1}{h(\phi)}-1}.
 \end{align}
\end{subequations}
These relations work when both $W_\phi$ and $W_\chi$ do not vanish. But we can also consider the case where $W_\chi=0$, which must be imposed that
\begin{subequations}
\label{f1}
 \begin{align}
  W_{\phi} &= \mathcal{W}_{\phi},\\
  f(\chi) &= \frac{1- \mathcal{W}_{\chi}}{\alpha}, \\
  g(\phi) &= \frac{1}{b\, h(\phi)}.
 \end{align}
\end{subequations}
The above results are both new and general, so we explore the corresponding correlations considering two distinct possibilities, one in which $W_{\chi} = 0$ and another where $W_{\chi}\neq 0$. We emphasize that this correspondence with geometrical constraints depends on the specific derivative interaction introduced in Eq. \eqref{lgeo}. While alternative Lorentz-breaking terms may also admit a similar correlation, this would depend on the detailed structure of the coupling and must be investigated individually.

Since our goal within this family of models is to demonstrate the possibility of mimicking the solutions of the Lorentz-invariant geometric constraint, we highlight only the features that differ from the Lorentz-invariant case.

\subsubsection{First model}
\begin{figure}[!ht]
    \centering
\includegraphics[width=0.9\columnwidth,height=5.2cm]{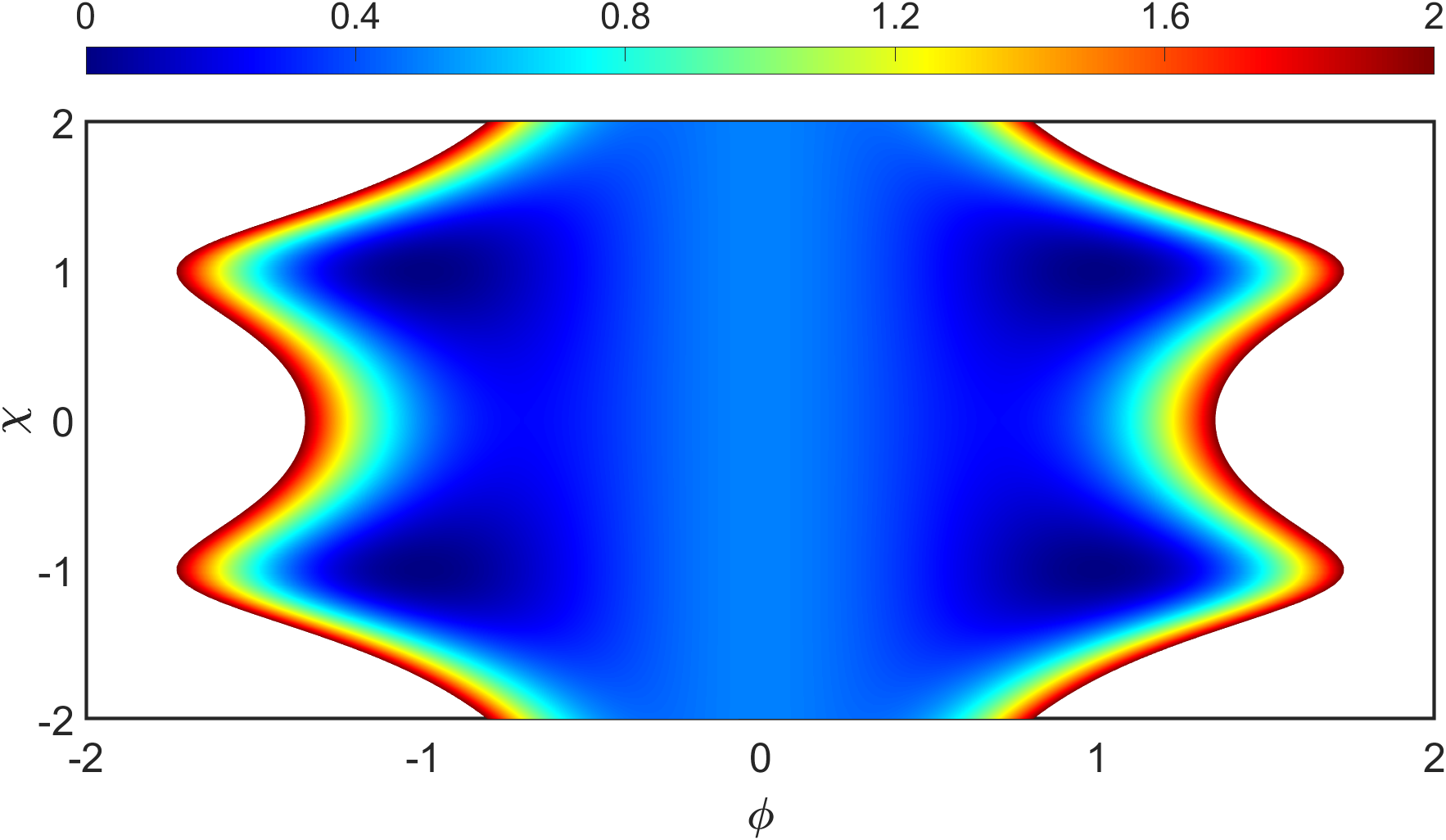}
    \caption{Top view of the potential \eqref{Pot1} in the $(\phi,\chi)$ plane.}
    \label{fig1}
\end{figure}

To illustrate the correspondence between Lorentz breaking and geometrically constrained solutions, let us first examine the case where $W_\chi=0$. Consider that the model whose solution we aim to reproduce has its auxiliary function defined as
\begin{equation}
\label{wcons}
    \mathcal{W}(\phi,\chi) = \phi - \frac{1}{3}\phi^3 + \chi - \frac{1}{3}\chi^3,
\end{equation}
indicating that both fields are of the fourth-order type. Furthermore, we have set the function $h(\phi)$ to be given by $h(\phi) = 1/\phi^2$. Since we consider $W_{\chi}=0$, the correspondence equations \eqref{f1} become
\begin{subequations}
 \begin{align}
  W_{\phi} &= 1-\phi^2,\\
  f(\chi) &= \frac{\chi^2}{\alpha}, \\
  g(\phi) &= \frac{\phi^2}{b}.
 \end{align}
\end{subequations}
Thus, the potential \eqref{vg} must take the form
\begin{equation}
\label{Pot1}
    V(\phi,\chi) = \frac{1}{2}\PC{1-\phi^2}^2 + \frac{1}{2}\phi^4\PC{1-\chi^2}^2,
\end{equation}
which has minima at $\phi_{\pm} = \pm 1$ and $\chi_{\pm} = \pm 1$. This potential is illustrated in the $(\phi,\chi)$ plane in Fig. \ref{fig1}.

Although we still have the same potential minima as the model discussed in \cite{geometrical}, which is mandatory to reproduce its solutions, the potential function itself is different. Moreover, the first-order equations \eqref{1op} are now written as
\begin{subequations}
 \begin{align}
  \phi' &= 1-\phi^2, \label{p4}\\
  \chi' &= \phi^2\PC{1-\chi^2},
 \end{align}
\end{subequations}
which are identical to the differential equations studied before in \cite{geometrical} for $\alpha =1$. Therefore, the solution is $\phi(x) = \tanh(x)$ and $\chi(x) = \tanh\PC{x-\tanh(x)}$, clearly displaying the geometric constraint related to magnetic materials \cite{geometrical,PRB}. Notice that in the definition of the functions $f(\chi)$ and $g(\phi)$ we have intentionally worked to eliminate the parameters $\alpha$ and $b$ from the potential and consequently from the first-order equations. This is important since these parameters play different roles in this model, when compared to the model discussed in Ref. \cite{geometrical}.

Beyond the potential, another distinction emerges when analyzing the energy density. Since $W_{\chi} = 0$, the field $\chi$ has no influence on $\rho$ for solutions that obey the first-order formalism, which means that $\rho(\phi,\chi) = \rho(\phi)$. So, the influence of the internal structure encoded in the solution $\chi(x)$ does not emerge on the energy density; consequently, its profile is not displayed.

\subsubsection{Second model}
\begin{figure}[!ht]
    \centering
    \includegraphics[width=0.9\columnwidth,height=5.2cm]{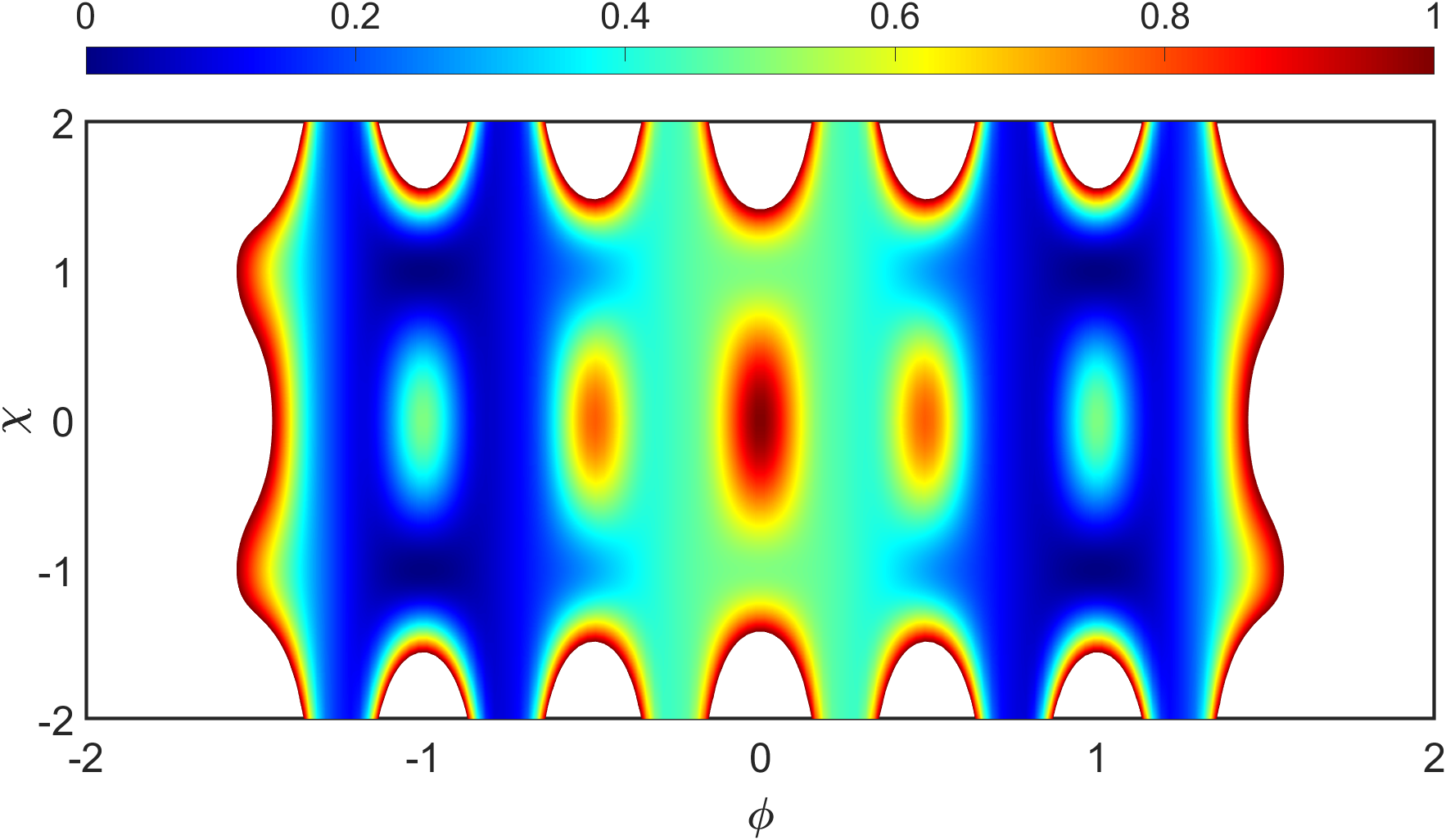}
    \caption{Top view of the potential \eqref{Pot2} in the $(\phi,\chi)$ plane, for $n=2$.}
    \label{fig2}
\end{figure}

 Now, let us deal with the case where $W_{\chi}\neq 0$. To analyze this scenario, we consider the same auxiliary function described in \eqref{wcons}, but now define the coupling function as $h(\phi)=1/\cos^2(n\pi\phi)$, with $n\in \mathbb{Z}$. The constraints \eqref{f2} are rewritten as

\begin{subequations}
 \begin{align}
  W_{\phi} &= 1-\phi^2, \\
  W_{\chi} &= 1-\chi^2,\\
   g(\phi) &= \frac{1}{b}\PC{\cos^2(n\pi\phi)-1}.
 \end{align}
\end{subequations}
The potential \eqref{vg} becomes
\begin{equation}
\label{Pot2}
    V(\phi,\chi) = \frac{1}{2}\PC{1-\phi^2}^2 + \frac{1}{2}\cos^4(n\pi\phi)\PC{1-\chi^2}^2,
\end{equation}
which has minima at $\phi_{\pm} = \pm 1$ and $\chi_{\pm} = \pm 1$. Its behavior is illustrated in Fig. \ref{fig2}. The potential \eqref{Pot2} contains two distinct contributions, one polynomial and the other periodic. Since the polynomial term increases rapidly, we established an upper limit for the color scale to highlight the minima of the potential. However, as a consequence, many of the associated periodic maxima are not seen in the figure.

As was pointed out in the previous model, the potential function above is different from the one discussed in \cite{geometrical}, although they have the same set of  minima. Since we are still working with a $\phi^4$ model, the first-order equation for $\phi$ remains $\phi' = 1-\phi^2$, with a nontrivial solution given by $\phi(x) = \tanh(x)$. The first-order equation for $\chi$ is given by
\begin{equation}
    \chi' = \cos^2(n\pi\phi)\PC{1-\chi^2},
\end{equation}
and its solution is
\begin{equation}
    \chi(x) = \tanh(\gamma(x)),
\end{equation}
where
\begin{equation}
    \gamma(x) = \frac{x}{2} + \frac{1}{4}\!\PC{\Ci(\xi_+) - \Ci(\xi_-)},
\end{equation}
and
\begin{equation}
    \Ci({\xi_\pm})=\Ci( 2n\pi(1\!\pm\!\tanh(x))),
\end{equation}
is the cosine integral function, as defined in \cite{Ci}. This solution matches the one found in \cite{geometrical} (when $\alpha=1$), for the model described by \eqref{lcons}, as expected. In addition to the case where $W_{\chi} = 0$, this new example also works to support the idea that the generalized Lorentz-breaking theory described by \eqref{lgeo} can reproduce solutions engendering geometric constraints, for both $W_{\chi} = 0$ or $W_{\chi}\neq 0$. Furthermore, since the solutions and auxiliary functions are the same as those discussed in \cite{geometrical}, the energy density of the solution is also the same.

Although the Lorentz-breaking scenario successfully reproduces the solutions of the Lorentz-invariant geometrically constrained framework, its applications go beyond this. In subsequent families of models, we uncover the emergence of novel structures.

\subsection{Second Family}

First, let us consider the case where the coupling of the fields only occurs through the Lorentz-breaking term and in which the functions $g(\phi)$ and $f(\chi)$ remain arbitrary. For simplicity, we suppose that the auxiliary function has the form

\begin{equation}
    W(\phi,\chi)= W(\phi).
\end{equation}
Despite the fact that this is a simple example, nontrivial solutions arise from it along with new features. Using equation \eqref{vg}, the potential becomes
\begin{equation}
\label{pot1}
    V(\phi,\chi) = \frac{1}{2}W_{\phi}^2+\frac{1}{2}b^2g^2(\phi)\PC{1-\alpha f(\chi)}^2,
\end{equation}
whose minima will only be defined when we specify all the functions. The first-order equations \eqref{primeira ordem de phi} and \eqref{primeira ordem de chi} take the form
\begin{subequations}\label{eq.1grau}
 \begin{align}
  \phi' &= W_{\phi},  \label{phi 1-1}\\
  \chi' &=  bg(\phi)\PC{1-\alpha f(\chi)} \label{chi 1-1}. 
 \end{align}
\end{subequations}
Since the auxiliary function $W$ does not depend on the field $\chi$, the equation \eqref{phi 1-1} can be solved independently. Its solution is denoted as $\phi(x) = \phi_{s}(x)$ and by substituting it in \eqref{chi 1-1}, one has 

\begin{equation}
    \chi' =  bg(\phi_{s}(x))\PC{1-\alpha f(\chi)} \label{chi1}.
\end{equation}
The existence of an analytical solution for this equation depends on the form of the functions $f(\chi)$ and $g(\phi)$, along with the analytic form of $\phi_s(x)$. Nevertheless, it is always possible to write $\chi(x) = \chi_{s}(\varepsilon(x))$, where 

\begin{equation}
    \varepsilon(x) = b \int g(\phi_{s}(x))\;dx.
\end{equation}

 Notice that $b$ is the spatial component of the vector $k^\mu$ that breaks the Lorentz symmetry, which modulates the strength of the interaction between the two fields. The above solution shows that the field $\phi$ induces a geometric constraint on the field $\chi$, with $\phi_{s}$ altering the behavior of the solution $\chi(x)$. Also, we were able to define a new coordinate $\epsilon(x)$ that parametrizes the first-order equation for $\chi$ in the form $d\chi/d\epsilon = \PC{1-\alpha f(\chi)}$. If we set $b=0$ the one field standard scenario is recovered, with no Lorentz breaking.

We also highlight that the geometric constraint observed in the above kink-like system arises from a different interaction when compared to the one previously investigated in \cite{geometrical}, where the constraint emerged through a coupling via a specific kinetic term. Moreover, it also differs from the models investigated in the first family as it does not require any specific choice for the function $f(\chi)$. 

Furthermore, the fact that $\chi(x)$ is a localized solution actually plays the interesting role of restricting the effect of a preferential direction to a finite region of space. This is a consequence of the already discussed boundary condition $\chi'(\pm \infty)\rightarrow0$, which means that the Lorentz-breaking term $bg(\phi_{s})\PC{1-\alpha f(\chi)}$ also goes to zero in the same limit, if $g(\phi_{s})$ and $f(\chi_{s})$ do not diverge. But still, the function $g(\phi)$ and the solution $\phi_s(x)$ can modify the internal structure of the $\chi$ field solution.

Another very important property that exists in this family lies on the definition of its energy density. Since $W(\phi)$ does not depend on $\chi$, the energy density can be written in the form
\begin{equation}
\label{ed1}
    \rho(\phi) = W_{\phi}\phi'.
\end{equation}
Thus, even though we can construct a model with the $\chi$ field having nontrivial behavior, if $\phi$ has a constant solution the energy density of our model remains zero. It is also important to remark that, since the field $\phi$ is not altered by the Lorentz-breaking term, any new feature of the system is encoded on the field $\chi$. In this sense, there will be no new behavior arising from $\rho$, so we do not display its profile for any model within this family.

\subsubsection{First model}
\begin{figure}[!ht]
    \centering
    \includegraphics[width=0.9\columnwidth,height=5.2cm]{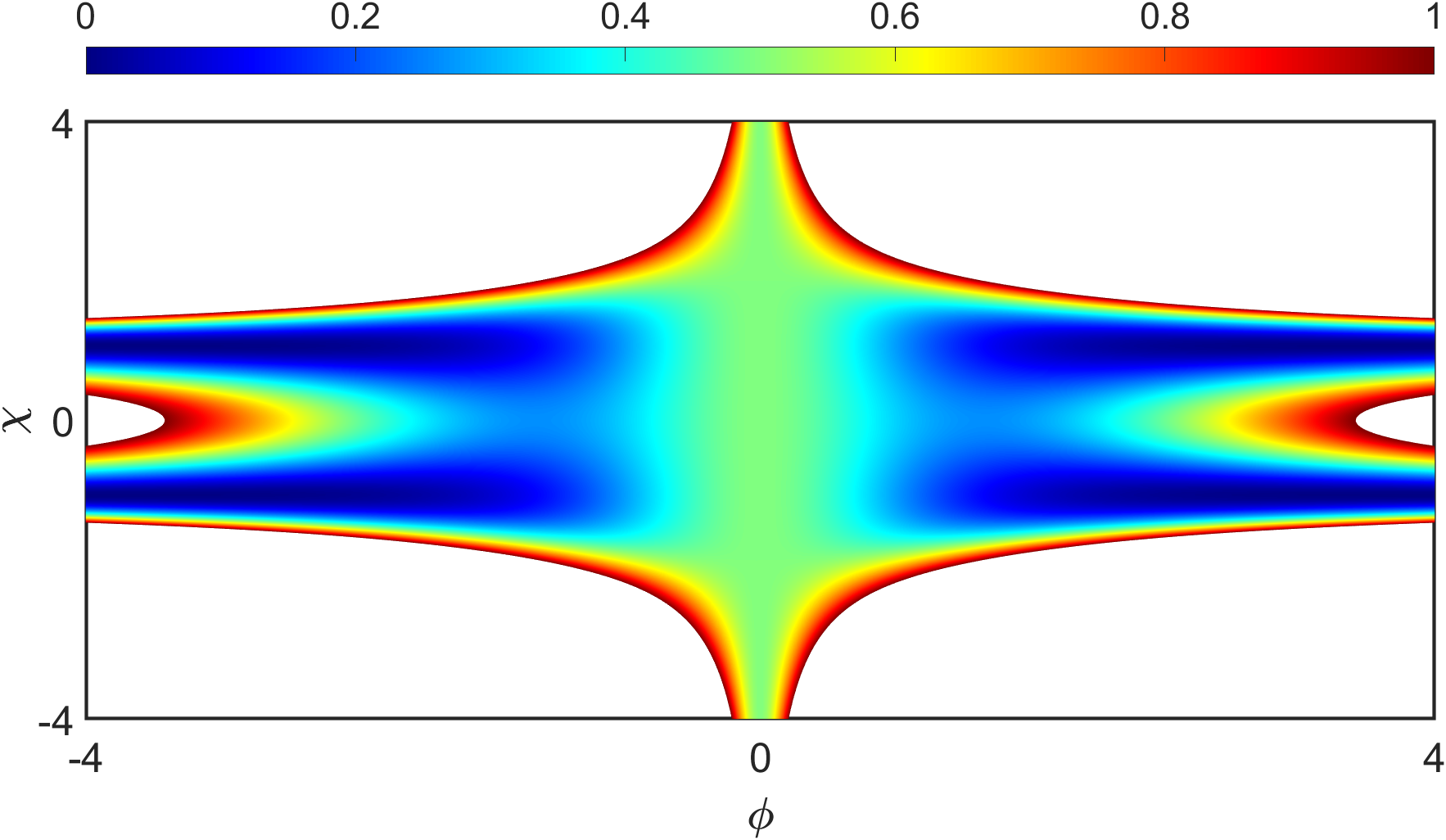}
    \caption{Top view of the potential \eqref{Pot3} in the $(\phi,\chi)$ plane, for $\alpha=1$ and $b=0.4$.}
    \label{fig3}
\end{figure}

As a first example, let us consider that $\phi$ is driven by a vacuumless model, which is described as in \cite{v1,v2} and references therein. The corresponding auxiliary function is defined as

\begin{equation}
    W(\phi) = \arctan[\sinh(\phi)].
\end{equation}
Selecting $f(\chi) = \chi^2$ and $g(\phi)=\phi$, the potential \eqref{pot1} is rewritten as 
\begin{equation}
\label{Pot3}
    V(\phi,\chi) = \frac{1}{2}\sech^2(\phi) + \frac{1}{2}b^2\phi^2(1-\alpha\chi^2)^2,
\end{equation}
which has its minima at $\phi_{\pm} \rightarrow \pm\infty$ and $\chi_{\pm}=\pm 1/\sqrt{\alpha}$. An illustration of this potential is displayed in Fig. \ref{fig3}. The fact that global minima occur asymptotically at infinite values of the field $\phi$ is clearly identified in the potential behavior.

The system of first-order equations takes the form
\begin{subequations}
\begin{align}
    \phi' &= \sech(\phi), \label{pvac}\\
    \chi' &= b\phi\PC{1-\alpha\chi^{2}}.
\end{align}
\end{subequations}
Notice that, if we set $g(\phi)=1$ the first-order equation for $\chi$ would mimic the one related to the $\chi^4$ model. Considering that both fields engender nontrivial behavior, one obtains the solution 

\begin{equation}
    \phi(x) = \arcsinh(x), \label{V} 
\end{equation}
and
\begin{equation}
\label{CV}
    \chi(x) = \frac{1}{\sqrt{\alpha}}\tanh\PC{\sqrt{\alpha}\;b\PC{x\arcsinh(x) - \sqrt{1+x^2}}}.
\end{equation}

\begin{figure}[!ht]
    \centering
    \includegraphics[width=0.9\columnwidth,height=5.2cm]{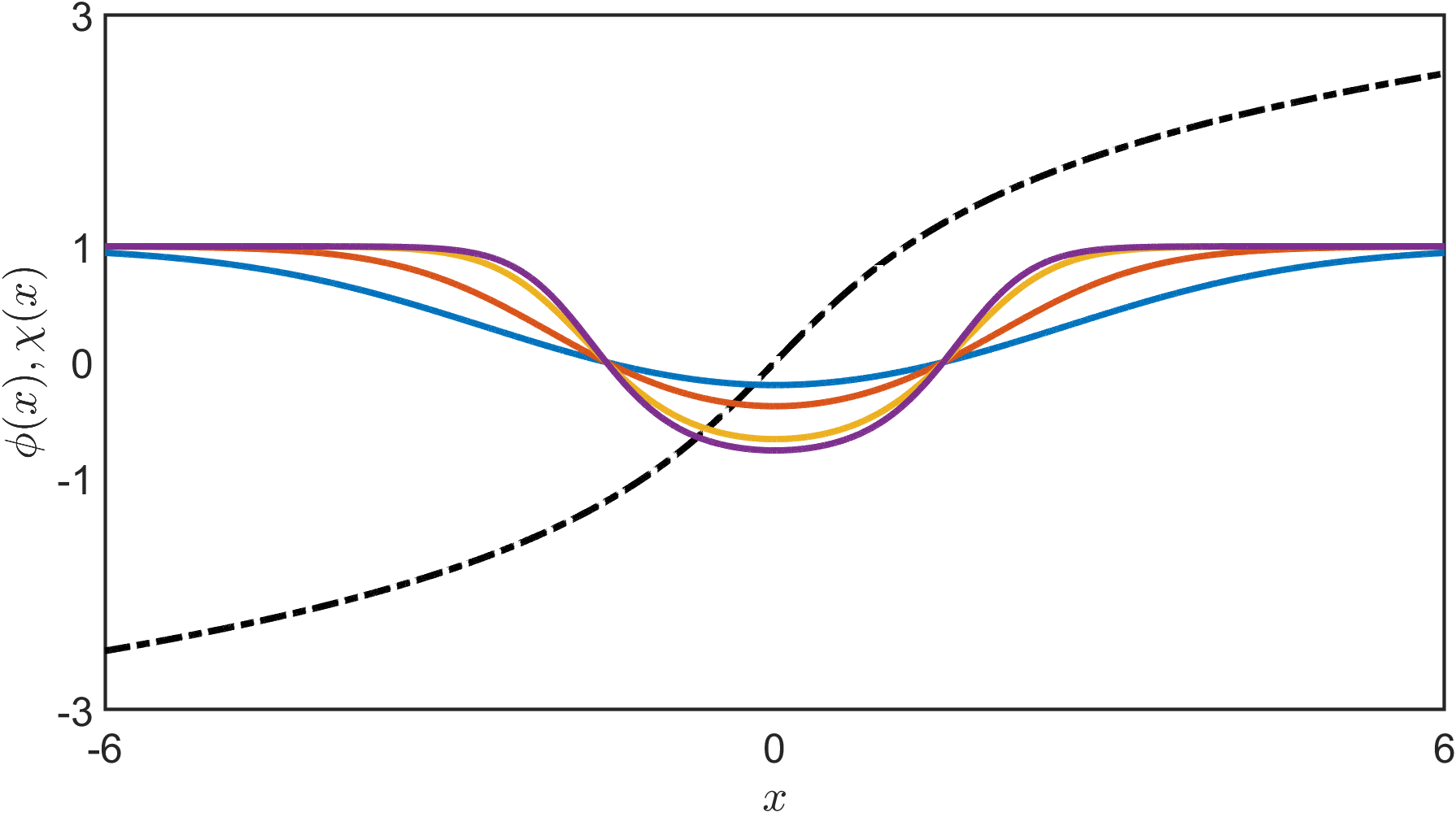}
    \includegraphics[width=0.9\columnwidth,height=5.2cm]{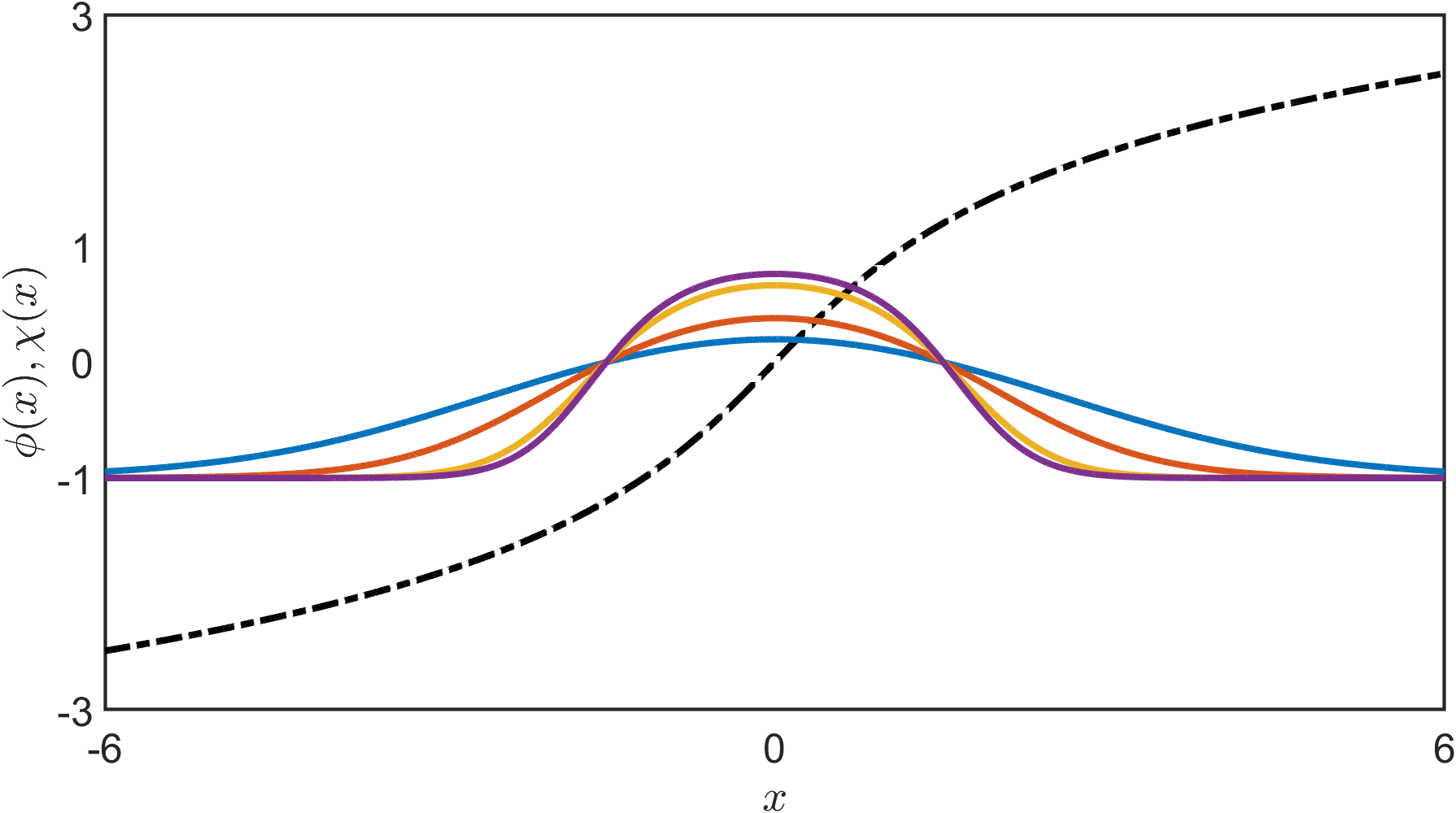}
    \caption{The solutions for $\phi(x)$ \eqref{V} and $\chi(x)$ \eqref{CV} are shown for $\alpha=1$, with the dash-dotted black line representing $\phi(x)$. We depict $\chi(x)$ for $b= \pm0.2, \pm0.4, \pm0.8, \pm1$, represented by blue, red, yellow and purple lines, respectively. The top and bottom panels refer to positive and negative values of $b$, respectively.}
    \label{fig4}
\end{figure}

Some solutions are depicted in Fig. \ref{fig4}. It is important to notice that the influence of the field $\phi$ over $\chi$ has induced a lump-like shape on $\chi(x)$, where the absolute value of the parameter
b alters the width of the solution and its sign is directly related to the concavity of the bell-shape profile. Models with two real scalar fields that admit such solutions were investigated before in systems with Lorentz breaking \cite{barreto} and without Lorentz breaking \cite{BSR,BNRT}. In both cases, however, the auxiliary function $W$ involved a coupling between the two fields.

The form of the solution $\chi(x)$ highly depends on the choice of the functions $g(\phi)$ and $f(\chi)$. The first, as mentioned before, can act on the formation of internal structures and the definition of the new coordinate $\epsilon(x)$, while the latter works to define an effective auxiliary function for the field $\chi$.

The energy density associated with this solution is given by
\begin{equation}
    \rho(x)=\sech^2(\arcsinh(x)) = \frac{1}{1+x^2}.
\end{equation}
As stressed before, since no new features arise from the previous energy density, we do not depict its behavior.

\subsubsection{Second model}

\begin{figure}[!ht]
    \centering
    \includegraphics[width=0.9\columnwidth,height=5.2cm]{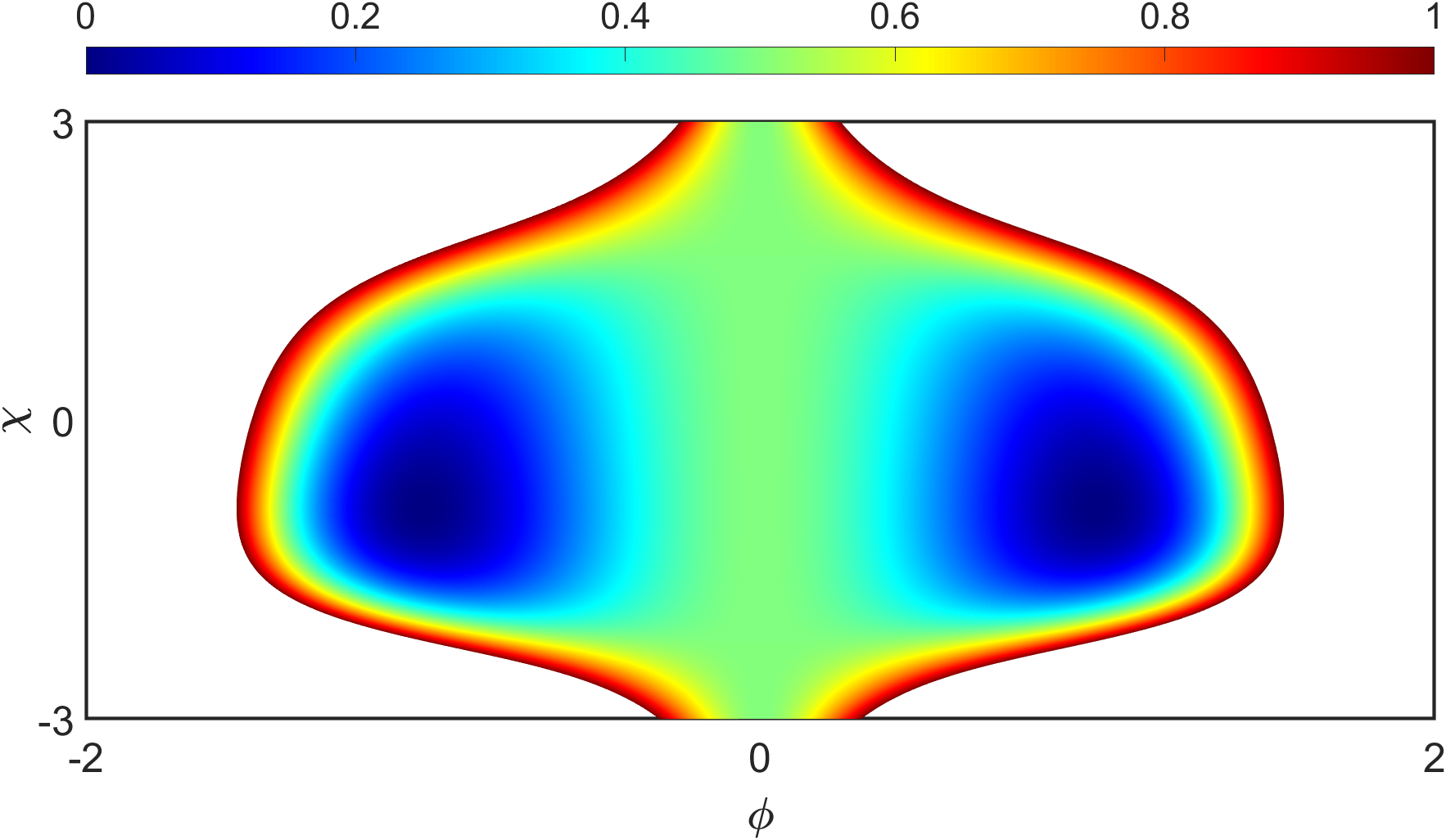}{\vspace{10pt}}
    \includegraphics[width=0.9\columnwidth,height=5.2cm]{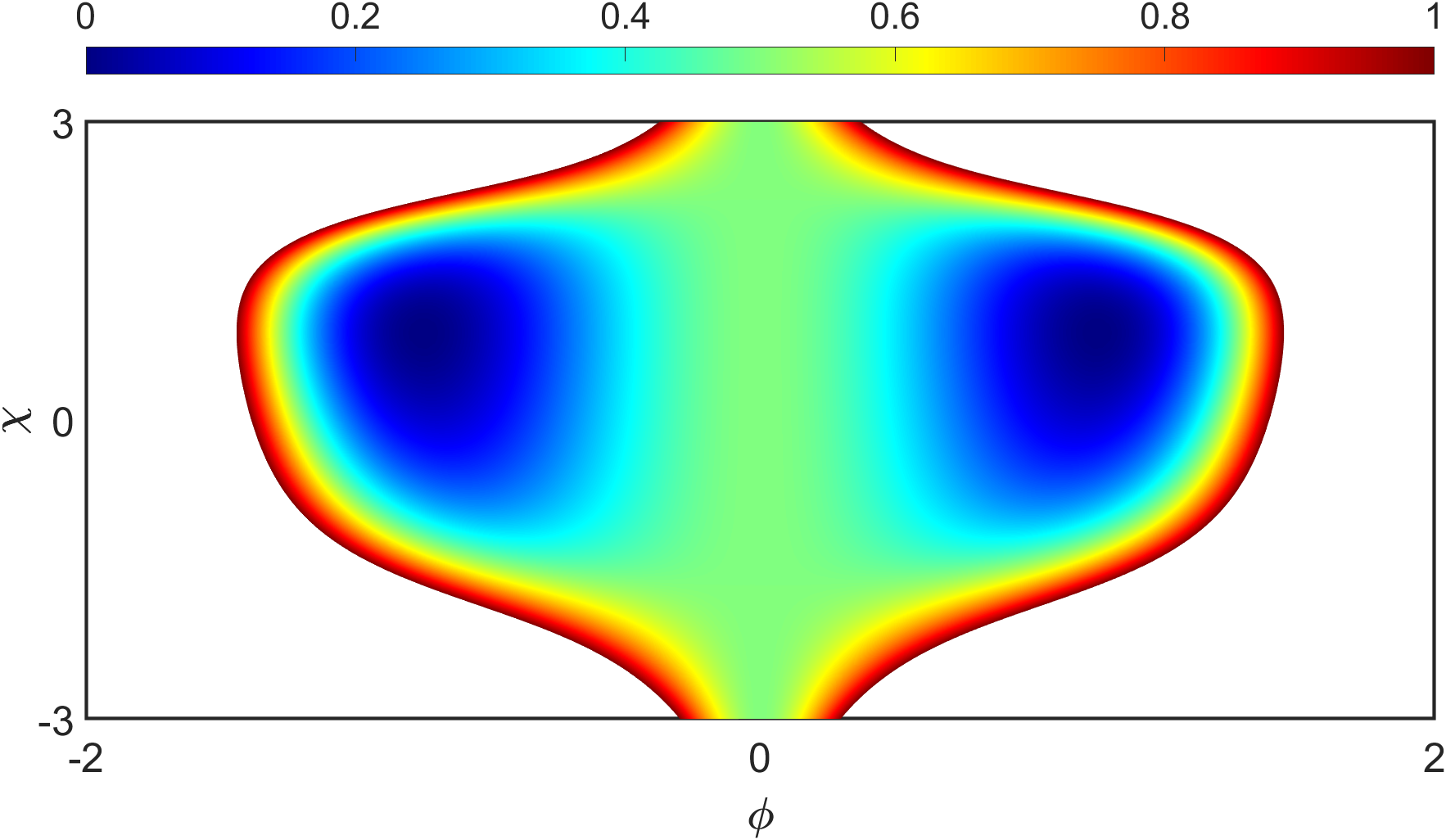}
    \caption{Top view of the potential \eqref{Pot4} in the $(\phi,\chi)$ plane, for $\alpha=-1$ and $b=-0.4$ (top) and for $\alpha=1$ and $b=0.4$ (bottom).}
    \label{fig5}
\end{figure}

There are several choices for $f(\chi)$ that lead to analytical kink-like solutions. Let us illustrate this by choosing $f(\chi)=\sinh(\chi)$ in the model where the auxiliary function is $W(\phi) = \phi - \phi^3/3$, while keeping $g(\phi)=\phi$. The potential \eqref{pot1} then becomes
\begin{equation}
\label{Pot4}
    V(\phi,\chi) = \frac{1}{2}\PC{1-\phi^2}^2+\frac{1}{2}b^2\phi^2\PC{1-\alpha\sinh(\chi)}^2.
\end{equation}
It has minima located at $\phi_0 = \pm1$ and $\chi_0 = \arcsinh(1/\alpha)$, which are clearly shown in the illustration depicted in Fig. \ref{fig5}, for different values of $\alpha$ and $b$.

Since the $\chi$ field has only one value associated with each potential minimum, any non-trivial solution for this field must be lump-like. As a consequence of considering a $\phi^4$ model, the first-order equation is the same as \eqref{p4} and can be solved independently to give $\phi(x) = \tanh(x)$. On the other hand, the first-order equation for $\chi$ takes the form

\begin{equation}
    \chi' = b\tanh(x)\PC{1-\alpha\sinh(\chi)},
\end{equation}
with solution
\begin{equation}
\label{cf2m2}
    \chi(x)= -2\arctanh\PC{\mathcal{P}(x)},
\end{equation}
where
\begin{equation}
\label{pf}
    \mathcal{P}(x) = \alpha - \sqrt{\alpha^2+1}\tanh\PC{\frac{b}{2}\sqrt{\alpha^2+1}\ln\PC{\cosh(x)}}.
\end{equation}

\begin{figure}[!ht]
    \centering
    \includegraphics[width=0.9\columnwidth,height=5.2cm]{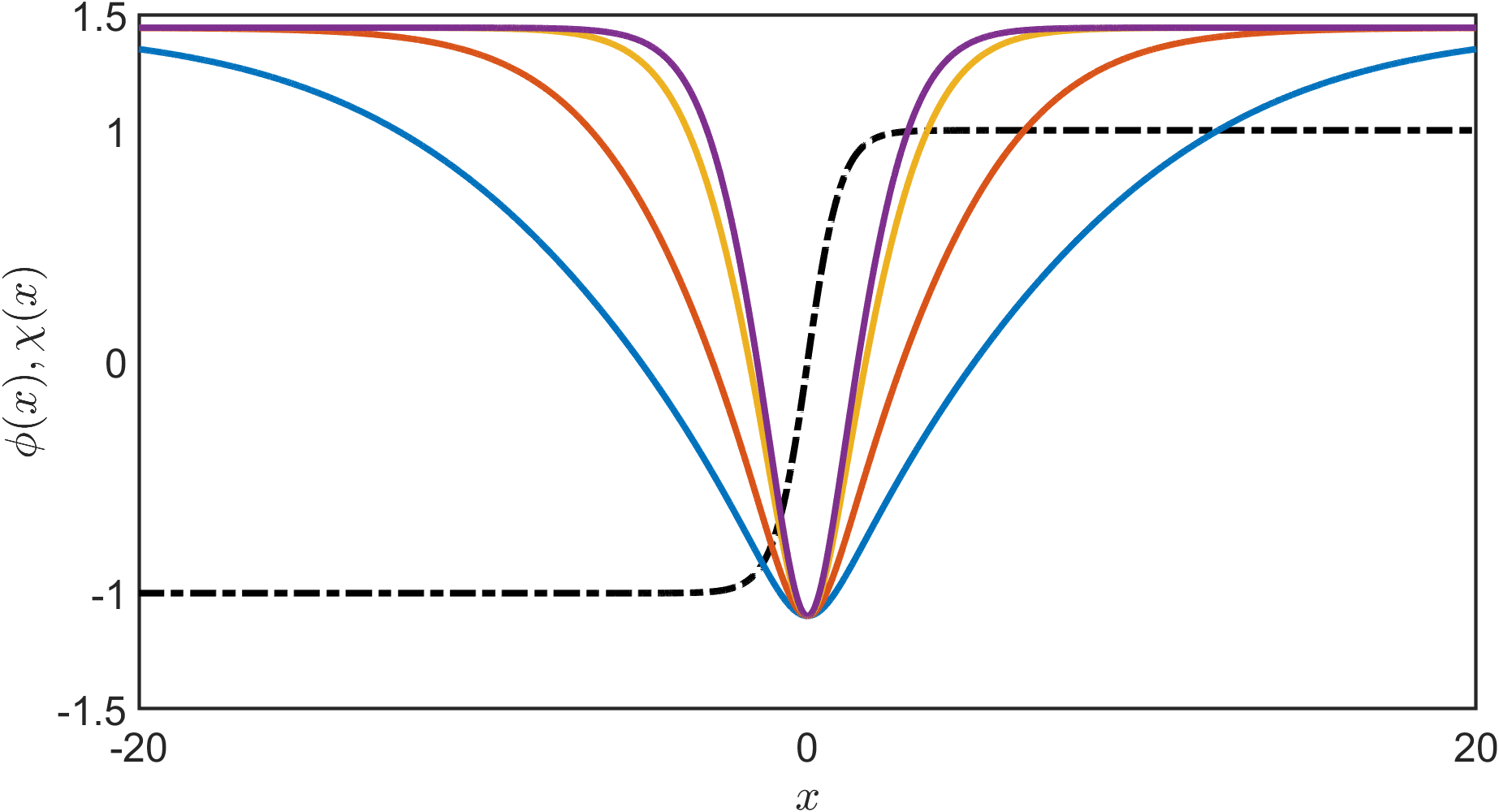}
    \includegraphics[width=0.9\columnwidth,height=5.2cm]{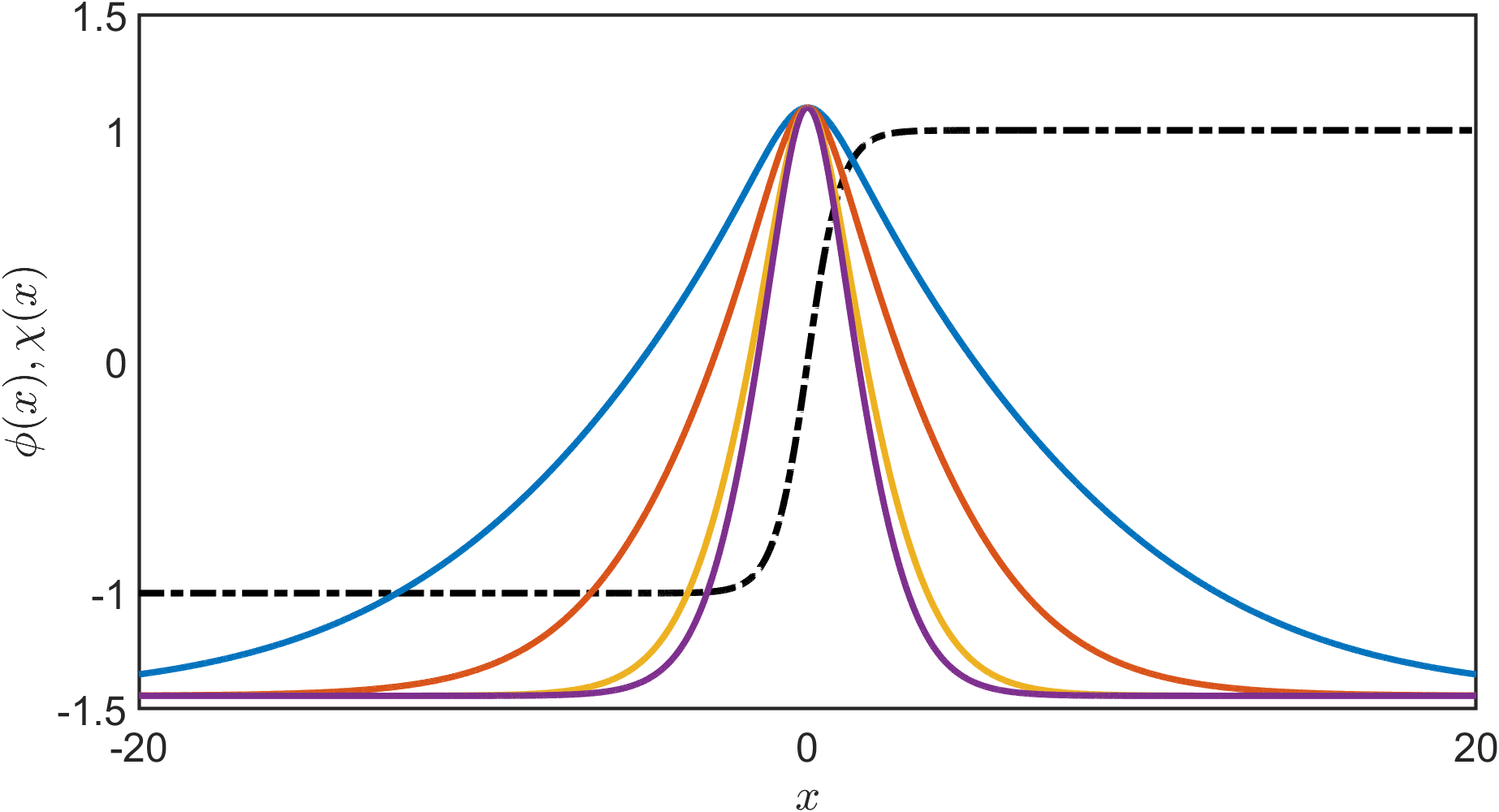}
    \caption{The solutions for $\phi(x)=\tanh(x)$ and $\chi(x)$ \eqref{cf2m2} are shown for $\alpha=0.5$ (top) and for $\alpha = -0.5$ (bottom), with the dash-dotted black line representing $\phi(x)$. We depict $\chi(x)$ for $b= \pm0.2, \pm0.4, \pm0.8, \pm1$, represented by blue, red, yellow and purple lines, respectively. The top and bottom panels refer to positive and negative values of $b$, respectively.}
    \label{fig6}
\end{figure}

It should be noticed that this solution is not defined for all values of $\alpha$, since it involves the function $\arctanh(x)$ whose domain is defined over the interval $\PR{-1,1}$. So, if $b$ is positive, one must consider $\alpha\in (0,1)$; if $b$ is negative, one must consider $\alpha\in (-1,0)$. The solution $\chi(x)$ is depicted in Fig. \ref{fig6} for several values of $b$. The lump-like behavior appeared in the $\chi$ solution as predicted before, with the parameter b controlling its width and concavity.

Since we are dealing with a $\phi^4$ model, the energy density is 

\begin{equation}
    \rho(x) = W_{\phi}\phi' = \sech^4(x).
\end{equation}

\subsection{Third Family}

 Let us now consider a family of models such that the auxiliary function depends on both fields but without coupling terms, i.e., $W(\phi,\chi) = \Gamma(\phi) + \Omega(\chi)$.  So, the interaction between the fields occurs only by the Lorentz-violating term, and in order to analyze general features
of this class of models, we do not define a specific form for any function. Thus, from equation \eqref{vg} we can rewrite the potential as

 \begin{equation}
 \label{potf3}
      V(\phi,\chi) = \frac{1}{2}\Gamma_{\phi}^2 +\frac{1}{2}\PC{\Omega_{\chi} + bg(\phi)\PC{1-\alpha f(\chi)}}^2,
 \end{equation}
 where $\Gamma_{\phi} = d\Gamma/d\phi$ and $\Omega_{\chi} = d\Omega/d\chi$. The first-order equations \eqref{primeira ordem de phi} and  \eqref{primeira ordem de chi} become

 \begin{subequations}
\label{1o2}
 \begin{align}
  \phi' &= \Gamma_{\phi},\label{primeira ordem phi familia 3}\\
  \chi' &= \Omega_{\chi} + bg(\phi)\PC{1-\alpha f(\chi)}\label{primeira ordem chi familia 3}.
 \end{align}
\end{subequations}

It should be noticed that the equation \eqref{primeira ordem phi familia 3}  can be independently solved since it only depends on the field $\phi$, and the solution is denoted by $\phi(x)=\phi_{s}(x)$. This result can be substituted into \eqref{primeira ordem chi familia 3} to give
\begin{equation}
    \chi' = \Omega_{\chi} + bg(\phi_{s})\PC{1-\alpha f(\chi)}.
\end{equation}

In order to express the geometric constraint explicitly, one can adopt a specific form for $f(\chi)$. By imposing the condition $\alpha f(\chi) = 1-\Omega_{\chi}$, one can factorize the $\chi$ dependence on the previous equation and get
\begin{equation}
\label{1ocf3}
    \chi' = \Omega_{\chi}\PC{1+bg(\phi_s)},
\end{equation}
which can be solved as
\begin{equation}
    \chi(x) = \chi_s(\gamma(x)), 
\end{equation}
where 

\begin{equation}
    \gamma(x) = x + b\int g(\phi_s(x))dx.
\end{equation}
The new coordinate $\gamma(x)$ responds as a geometric constraint imposed by the solution $\phi_{s}$ in the field $\chi$, and parametrizes the first-order equation \eqref{1ocf3} in the form $d\chi/d\epsilon = \Omega_{\chi}$. As previously emphasized, the parameter $b$ governs the strength of the coupling between the fields within the Lorentz-violating term. Setting $b=0$, the standard decoupled scenario is retrieved. 

The energy density associated with this model is given by  

\begin{equation}
    \rho = \Gamma_{\phi}\,\phi' + \Omega_{\chi}\,\chi'.\label{rhof2}
\end{equation}
Since the auxiliary function $W$ has contributions from both fields, its energy density will depend on the behavior of both $\phi(x)$ and $\chi(x)$. As a consequence, interesting features arise in the investigation of the energy density of such models, as we further explore below.

\subsubsection{An illustrative model}

\begin{figure}[!ht]
    \centering
    \includegraphics[width=0.9\columnwidth,height=5.2cm]{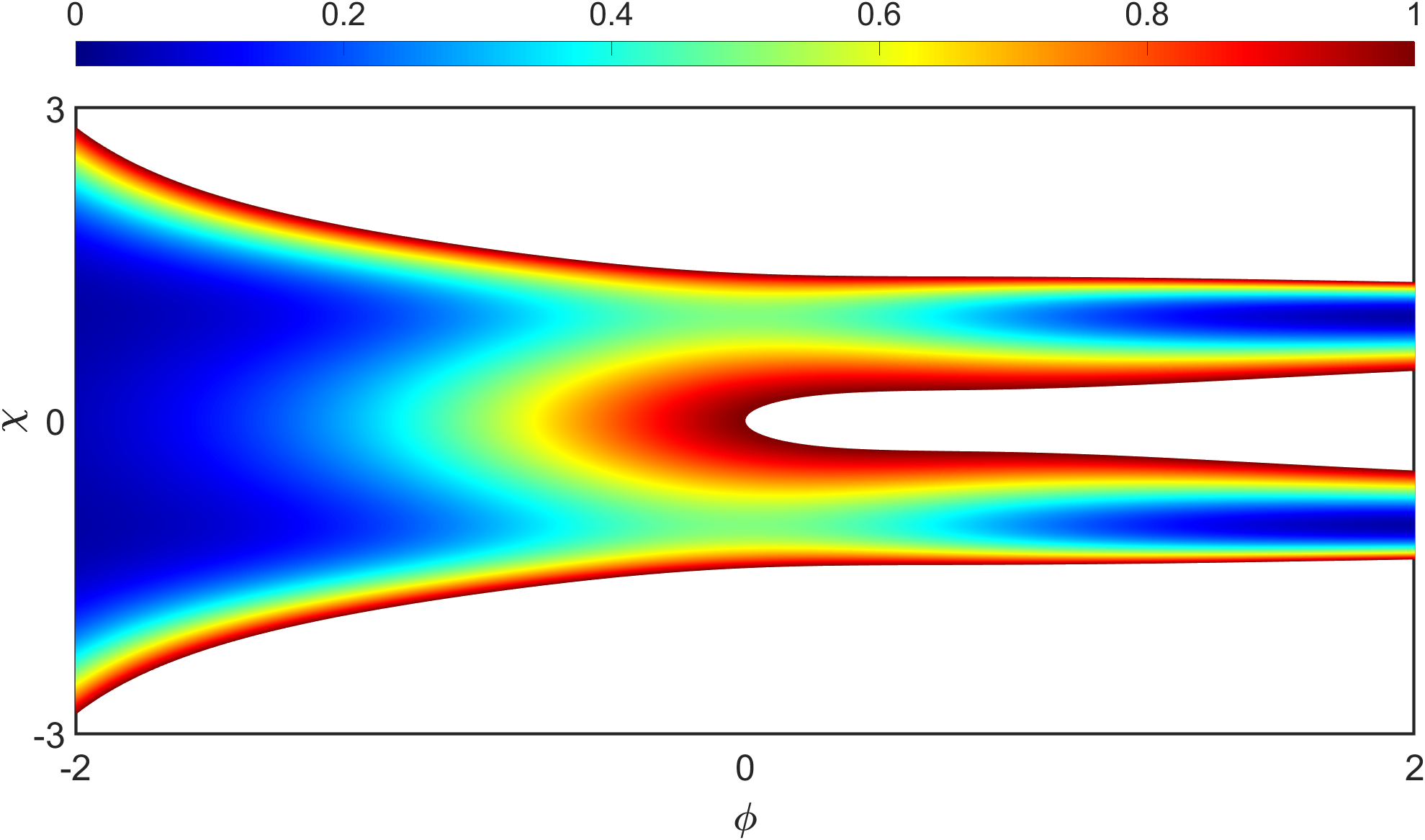}{\vspace{10pt}}
    \includegraphics[width=0.9\columnwidth,height=5.2cm]{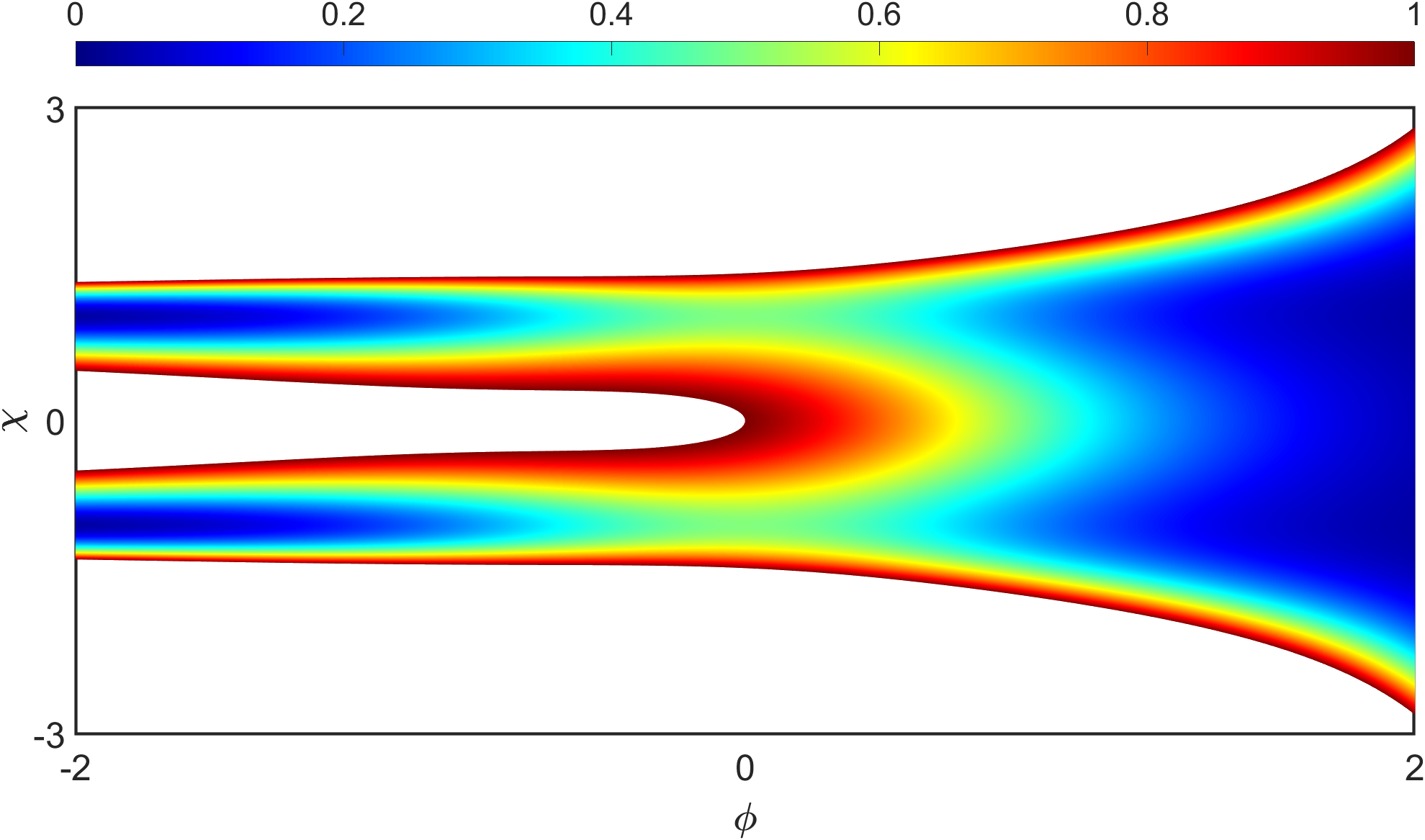}
    \caption{Top view of the potential \eqref{Pot5} in the $(\phi,\chi)$ plane, for $b=-0.4$ (top) and $b=0.4$ (bottom).}
    \label{fig7}
\end{figure}

In order to analyze how the geometric constraint acts in specific scenarios, let us consider a model where the auxiliary function terms are defined as

\begin{equation}
    \Gamma(\phi) = \arctan(\sinh(\phi)), \;\;\;\Omega(\chi)=\chi -\frac{1}{3}\chi^3.
\end{equation}
As mentioned above, we take a specific choice for $f(\chi)$, which for this model is $\alpha f(\chi) = 1-\Omega_{\chi} = \chi^2$. Assuming $g(\phi)=\phi$, the potential \eqref{potf3} becomes

\begin{equation}
\label{Pot5}
   V(\phi,\chi) = \frac{1}{2}\sech^2(\phi) +\frac{1}{2}\PC{1-\chi^2}^2\PC{1+b\phi}^2,
\end{equation}
which has minima located at $\phi_{\pm}=\pm\infty$ and $\chi_{\pm}=\pm1$. This potential is illustrated in Fig. \ref{fig7}, indicating that the minima are associated with infinite values of the field $\phi$. The asymmetry under inversion $\phi \rightarrow -\phi$ becomes evident due to the presence of the multiplicative factor $(1+b\phi)^2$ in the second term of the potential.

The first-order equations are

\begin{subequations}
 \begin{align}
  \phi' &= \sech(\phi),  \\
  \chi' &= \PC{1-\chi^2}\PC{1+b\phi}.\label{cv4}
 \end{align}
\end{subequations}
The differential equation for the field $\phi$ is the same found within the vacuumless model \eqref{pvac}, with the solution being $\phi(x)=\arcsinh(x)$. Now it is straightforward to substitute this expression in \eqref{cv4} to obtain the result

\begin{equation}
\label{cf3}
    \chi(x) = \tanh\PC{\gamma(x)} ,
\end{equation}
where
\begin{equation}
    \gamma(x) = x + b\PC{x\arcsinh(x) - \sqrt{x^2+1}}.
\end{equation}

\begin{figure}[!ht]
    \centering
    \includegraphics[width=0.9\columnwidth,height=5.2cm]{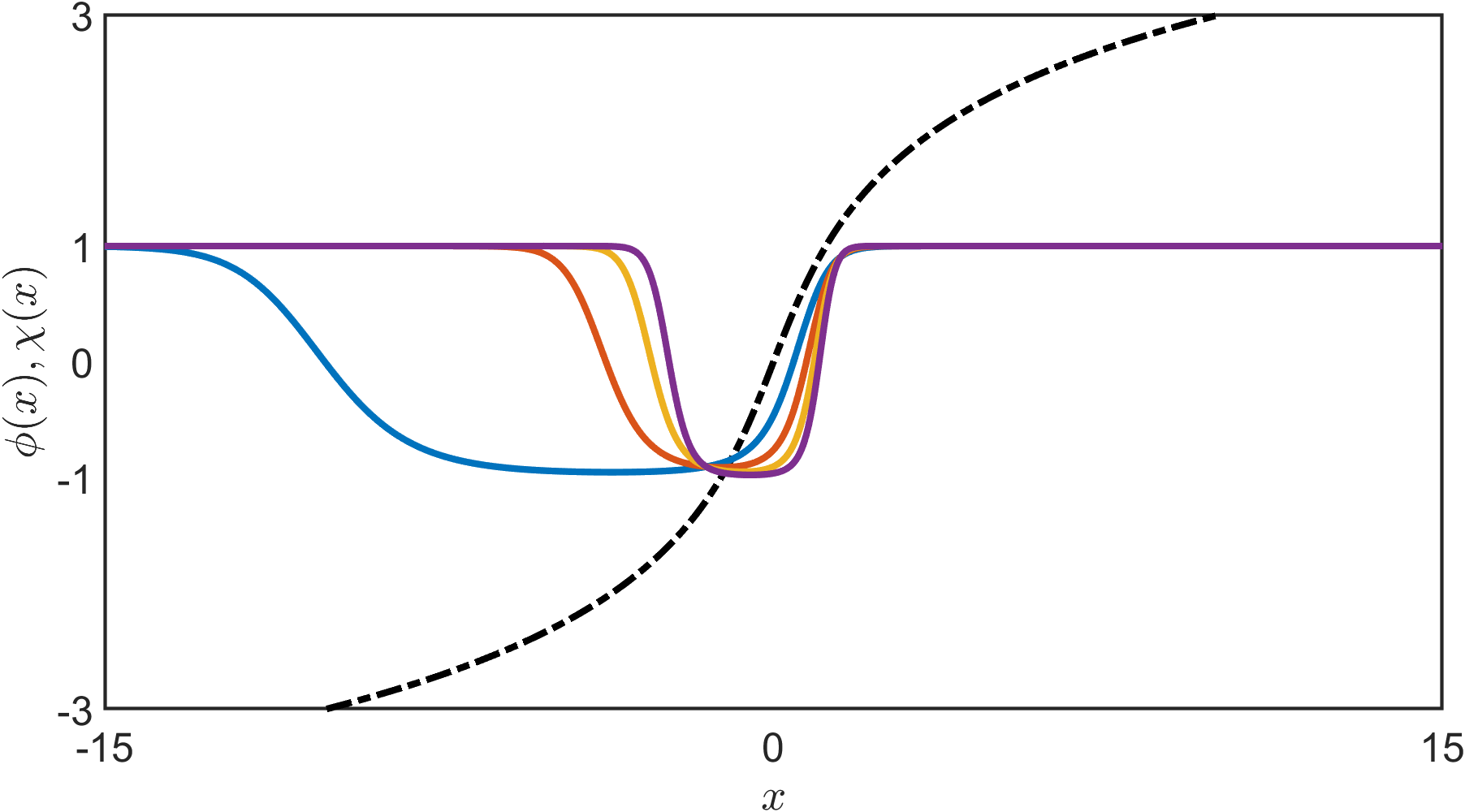}
    \includegraphics[width=0.9\columnwidth,height=5.2cm]{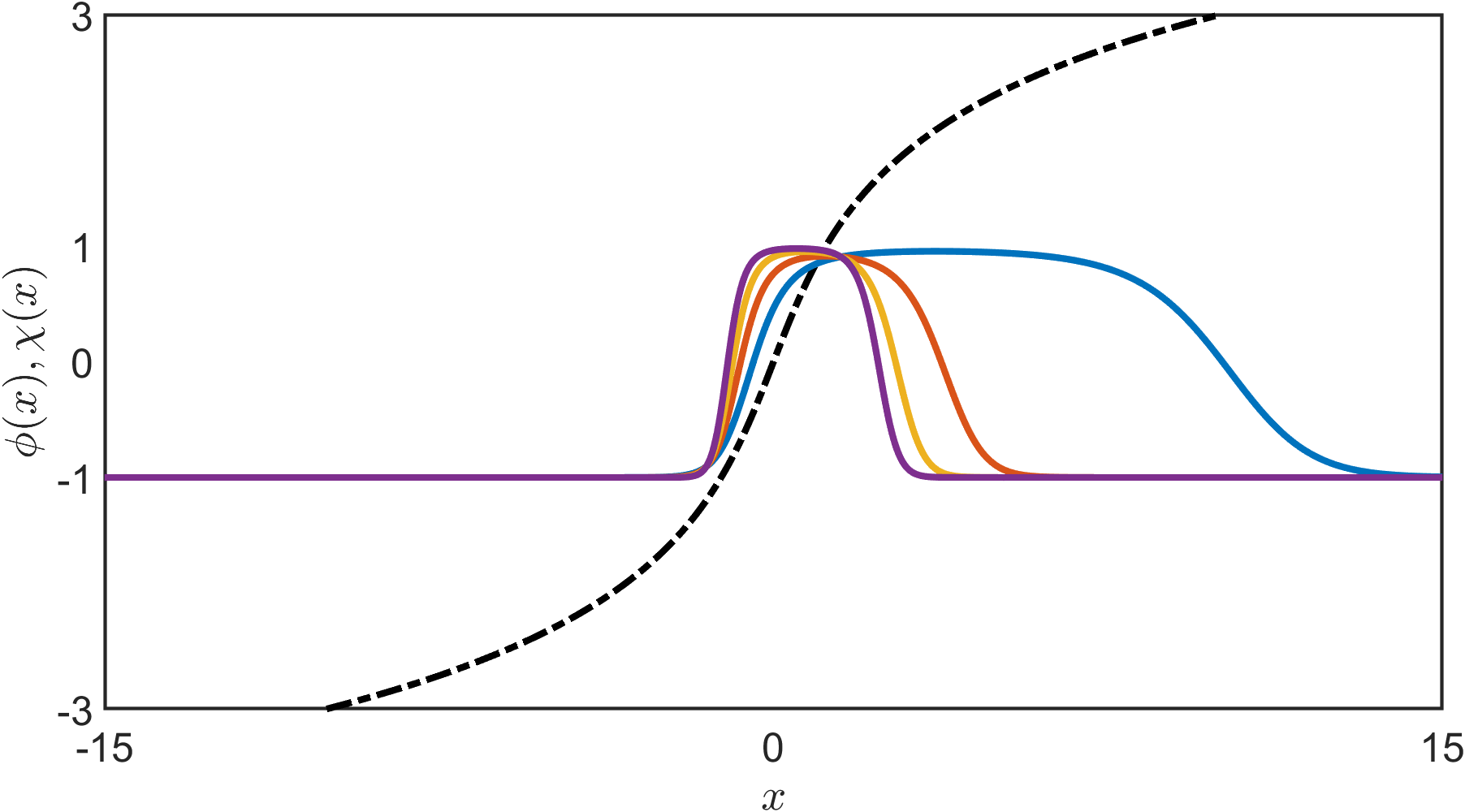}
    \caption{The solutions for $\phi(x)=\arcsinh(x)$ and $\chi(x)$ \eqref{cf3} are shown, with the dash-dotted black line representing $\phi(x)$. We depict $\chi(x)$ for $b= \pm0.5, \pm1, \pm1.5, \pm2$, represented by blue, red, yellow and purple lines, respectively. The top and bottom panels refer to positive and negative values of $b$, respectively.}
    \label{fig8}
\end{figure}

The shape of the solutions is depicted in Fig. \ref{fig8}. The $\chi(x)$ profile is lump-like for any $b \neq 0$, with the sign of $b$ related to the concavity of the bell shape. Still, one can note that the Lorentz parameter modulates the width of the solution, which decreases as $\vert b \vert$ increases. 

\begin{figure}[!ht]
    \centering
    \includegraphics[width=0.9\columnwidth,height=5.2cm]{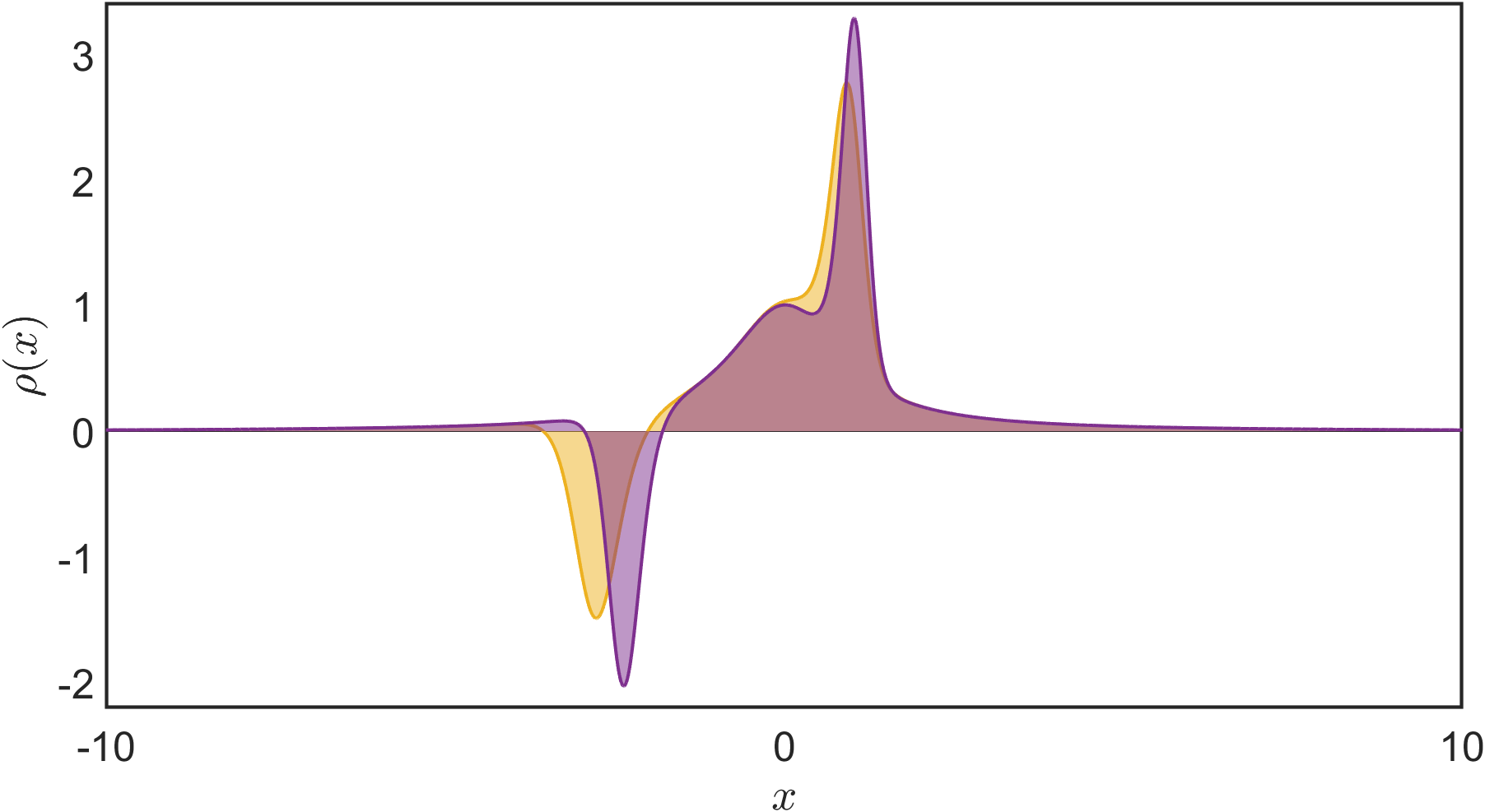}
    \includegraphics[width=0.9\columnwidth,height=5.2cm]{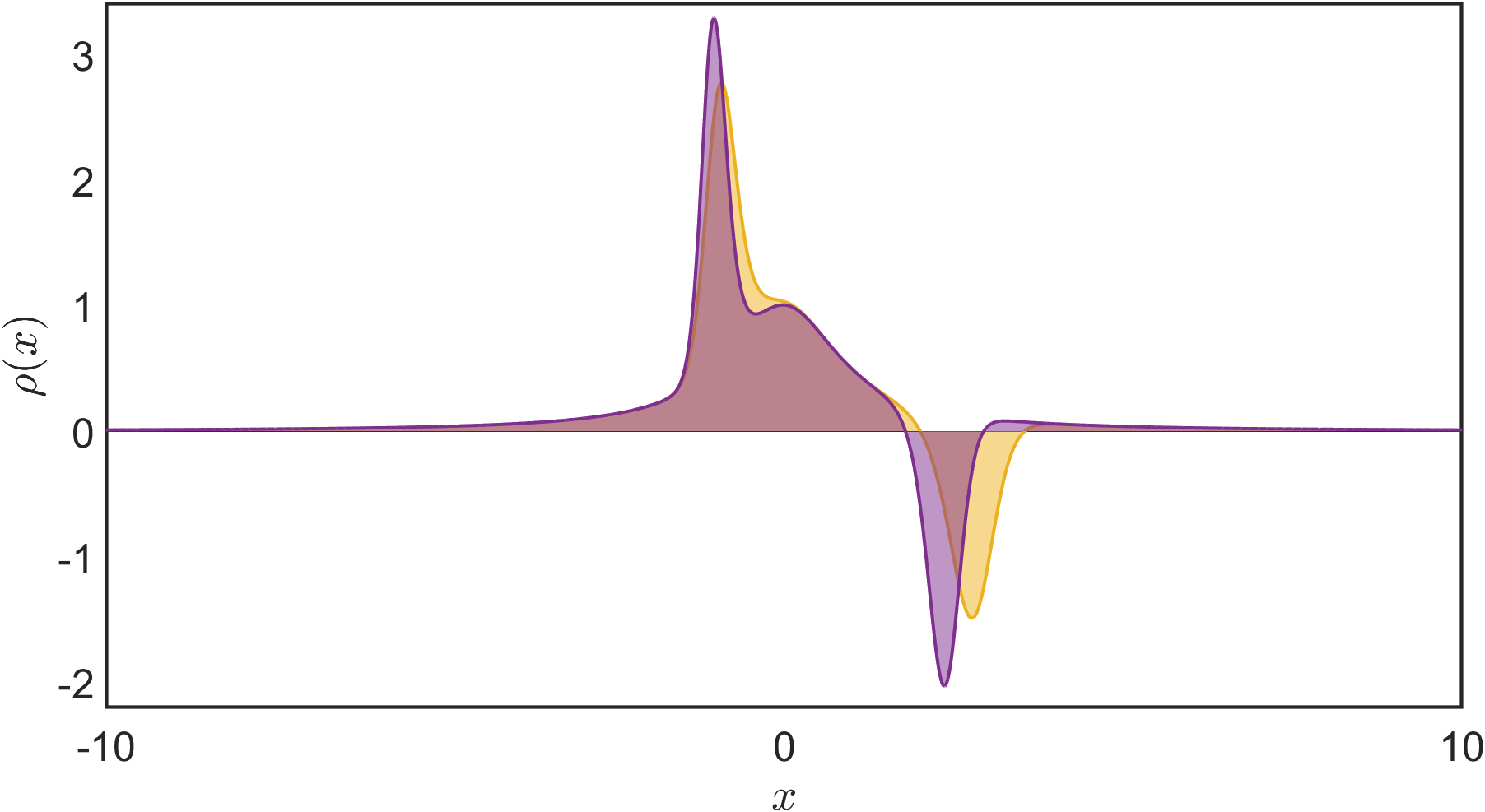}
    \caption{Energy density $\rho(x)$ \eqref{edf3}. We compare the situations where $b= \pm1.5, \pm2$ represented by yellow and purple areas, respectively. The top and bottom panels refer to positive and negative values of $b$, respectively. }
    \label{fig9}
\end{figure}

The energy density \eqref{rhof2} is now given by

\begin{equation}
\label{edf3}
    \rho(x) = \frac{1}{x^2+1} +
    \sech^4\PC{\gamma(x)}(1+ b\arcsinh(x)).
\end{equation}
Since $\chi(x)$ has a nonmonotonic behavior whenever $b\neq 0$, we have that $\rho(x)$ will also display regions with negative values within this interval of $b$. An illustration of the energy density is shown in Fig. \ref{fig9}. We further remark that in Ref. \cite{barreto} the investigation also found solutions engendering regions of negative energy density, but the stability analysis included there showed linear stability.  In this sense, the presence of regions of negative energy density does not necessarily imply the existence of instabilities of the localized solutions. It should be noted that solition solutions in the presence of a negative energy contribution have been considered many times before in various contexts, see e.g. \cite{Verbin:2007fa,Bechmann:1995sa,Ferreira:2025xey,Amari:2024pnw}. On the other hand,  
we have just seen that the energy densities of the two previous families of models do not present regions with negative values, so we cannot suggest that regions of negative energy density appear as a direct consequence of Lorentz violation.

%%%%%%%%%%%%%%%%%%%%%%%%%%%%%%%%%%%%%%%%%%%%%%%%%%%%%%%%%

%%%%%%%%%%%%%%%%%%%%%%%%%%%%%%%%%%%%%%%%%%%%%%%%%%%%%%%%%

\section{Conclusion}
\label{Con}
In this work, we have investigated distinct families of models, described by two real scalar fields in ($1+1$) spacetime dimensions. These models are constructed to break the Lorentz symmetry. They include an extra term in the Lagrangian, controlled by a constant vector that explicitly induces an anisotropy in the system, which works to violate Lorentz symmetry. The inclusion of this term is motivated by two previous investigations, one described in \cite{barreto}, exploring how the Lorentz breaking contributes to change the field configuration profiles, in comparison with the Lorentz-invariant case. The second investigation, a more recent study \cite{adam}, examines the phenomenon of a spectral wall in models that develop Lorentz breaking, with particular emphasis on connection with singularities of the dynamical vibrational moduli space. 

Throughout our investigation, a direct connection was established between Lorentz violation and the geometric constraint that was previously investigated in \cite{geometrical}, correlated with the experimental study of magnetization in constrained geometries \cite{PRB}. This novel correspondence may open promising directions for future applications. A possibility of current interest could be to investigate collisions between the new analytical solutions found in this work, to see how the presence of Lorentz breaking and the geometric constraint may modify the fractal structure previously identified in the scattering studied in Ref. \cite{Ya}. Indeed, we observe that some of the models previously discussed support bound modes within their stability spectra. Hence, the resonance energy transfer mechanism may also act on the non-relativistic collisional dynamics of the corresponding solutions. In this context, we think that a collective coordinate model approach can be employed to analyze how the presence of Lorentz breaking may influence the scattering structure; see, e.g., Refs. \cite{cc1,cc2,cc3,cc4} and references therein.

The first family introduced in the present work demonstrates that the solutions of a specific class of Lorentz-invariant models exhibiting geometric constraint \cite{geometrical} can be reproduced in a Lorentz-breaking framework, provided a specific choice of the functions within the model. The following families expand our analyses to new features, including a bell-shaped solution in a model with a separable auxiliary function (with $W(\phi,\chi)$ such that $W_{\phi\chi}=0$) or the appearance of geometrically constrained solutions associated with an energy density with regions having negative values. Interestingly, the study developed in Ref. \cite{adam} connects Lorentz violation to the spectral wall phenomenon \cite{adamspe}, so we think it may be of current interest to investigate the possibility of linking the spectral wall with constrained geometries.
It is also of current interest to investigate extended models involving additional scalar fields, in a way similar to the study considered in \cite{physd}, with three-field models in the presence of geometric constraints. A natural extension would be to incorporate Lorentz-violating terms into such frameworks, which seems to have no insuperable obstacle. Furthermore, one could consider alternative forms of Lorentz symmetry breaking, as the one considered before in Ref. \cite{baz01}, to examine whether a connection with geometric constraint is still possible. 

Scalar fields have been used to describe dark energy in cosmology, as investigated in \cite{Amen,dark}. In this context, interesting lines of further study can be related to domain walls \cite{dark} and also to the case of time crystals considered in \cite{Wil} and further examined in \cite{AA,Pinto,RCrA,RCrB} and references therein. Some of the above possibilities are now under consideration, and we hope to report the new results in a separate investigation. 

\acknowledgments{This work is supported by the Brazilian agencies Conselho Nacional de Desenvolvimento Cient\'ifico e Tecnol\'ogico (CNPq), grants Nos. 402830/2023-7 (DB and YS) and 303469/2019-6 (DB), and Coordenação de Aperfeiçoamento de Pessoal de Nível Superior (CAPES), Grants No. 88887.947425/2024-00 (GHB) and No. 88887.132514/2025-00 (GSS)}.

{\textbf {Data availability statement:}} 
All data that support the findings of this study are included within the article(and any supplementary files).

{\textbf{Authors statement:}} Authors have no relevant financial or non-financial interests to disclose.

{\textbf {ORCID IDs:}}

G. H. Bandeira: 0009-0006-1086-5990;

D. Bazeia: 0000-0003-1335-3705;

G. S. Santiago: 0009-0006-0529-4708;

Ya. Shnir: 0000-0002-9159-2675.

%\bibliographystyle{apsrev4-1} % Tell bibtex which bibliography style to use
%\bibliography{bibliography} % Tell bibtex which .bib file to use (this one is some example file in TexLive's file tree)

\end{document}